\documentclass[letterpaper, 10pt]{article}
\usepackage{preprint}

\title{Open-Source BOS Tomography Dataset of High-Speed Flow \\ Over a Flight Body}

\author{
  Joseph P. Molnar$^1$, Amit K. Singh$^1$,
  Christopher J. Clifford$^2$, Jordan D. Thayer$^2$,\\ Scott J. Peltier$^2$, Garrett C. Jones$^3$, and Samuel J. Grauer$^{1,4,}$\thanks{Corresponding author: \href{mailto:sgrauer@psu.edu}{sgrauer@psu.edu}}\vspace*{.3em}\\
  {\small $^1$Department of Mechanical Engineering, Pennsylvania State University}\vspace*{-.15em}\\
  {\small $^2$Aerospace Systems Directorate, Air Force Research Laboratory}\vspace*{-.15em}\\
  {\small $^3$Canvas, Inc.}\vspace*{-.15em}\\
  {\small $^4$School of Mechanical Engineering, Purdue University}\vspace{-1em}}
\date{}

\begin{document}

\maketitle
\setcounter{footnote}{3}
\vspace*{-2em}

\begin{abstract}
    \noindent We present an open-source background-oriented schlieren dataset with 70 views of high-speed flow over a flight body. Sample analyses are performed using a neural-implicit reconstruction technique (NIRT) with total variation regularization as well as data assimilation via the 3D compressible Euler equations. Limited-data reconstructions based on nine views resolve sharp shocks that are consistent with the geometry, reproduce validation deflections with high fidelity, and exhibit minimal artifacts. Data assimilation recovers unmeasured fields, marking the first demonstration of 3D state estimation directly from experimental schlieren measurements. The NIRT also enables efficient uncertainty quantification, providing insight into well-resolved flow features and guiding design-of-experiments efforts. Public access to the data and code repositories is detailed at the end of this correspondence. \par\vspace{.5em}
    
\end{abstract}
\vspace*{2em}

\section{Introduction}
\label{sec: introduction}
Background-oriented schlieren (BOS) tomography is a non-intrusive, low-cost optical diagnostic that provides quantitative, spatially resolved measurements of the refractive index field, which can be converted under suitable conditions to density, temperature, or species \cite{Raffel2015, Schmidt2025}. Each BOS image encodes a two-component deflection field resulting from line-of-sight integration of refractive index gradients. When multiple views are available, or a single view of an axisymmetric or planar flow, components of the gradient field may be independently reconstructed and then integrated using a Poisson solver to recover the refractive index field. In practice, density is often directly reconstructed by assuming uniform composition, i.e., a constant Gladstone--Dale coefficient. These steps may be performed sequentially or combined into a unified workflow, and some methods also recover velocity and pressure (energy, temperature, etc.) via data assimilation \cite{Molnar2023}.\par

Most BOS studies utilize single-camera measurements of axisymmetric flows \cite{Schmidt2025}, while engineering applications typically involve fully-3D fields. Reconstructing such flows requires multi-view imaging, with error inversely proportional to the number of views as independent views are added. In practice, the number of perspectives on offer is usually limited, especially for time-resolved measurements, where 5--15 views is typical \cite{Ota2011, Sourgen2012, Bathel2022, Takahashi2024, Yamagishi2025}. Accurate reconstruction in this regime requires prior information, such as flow symmetries, boundary conditions, or smoothness constraints. To advance research in this area, we have produced an open-source dataset featuring BOS measurements of high-speed flow over a model flight vehicle. The data include reference and deflected images from 70 perspectives, calibration data, deflection fields, and sample tomographic reconstructions. Along with these data, we include code for a neural-implicit reconstruction technique (NIRT) algorithm that is tailored for BOS tomography \cite{Molnar2025}. These resources are intended to support the development and benchmarking of BOS workflows, including deflection sensing, tomographic reconstruction, and data assimilation algorithms for compressible flows.\par

Several research groups have addressed the challenge of limited view counts in time-averaged BOS tomography for flight bodies by mounting the test article on a rotational strut and acquiring images successive angles. This ``rotisserie-style'' approach enables the reconstruction of asymmetric, 3D flows with a single camera, reducing costs and simplifying the optical setup. Notable examples include Takahashi et al. \cite{Takahashi2024} and Sourgen et al. \cite{Sourgen2012}, who recorded 90 and 19 perspectives, respectively, of various aerospace-relevant test articles. In both cases, reconstructed shock structures were broadly consistent with simulations but exhibited smearing artifacts, likely attributable to uncertainties in alignment and limitations of the filtered backprojection algorithm. Ota et al. \cite{Ota2011} used a similar mounting configuration to measure flow over an asymmetric cone with a telecentric BOS system. They achieved higher fidelity estimates by reconstructing the flow using an algebraic reconstruction technique, but the reconstructions still included non-physical features, such as an expansion upstream of the leading shock.\par

These demonstrations show that multi-view, time-averaged BOS can capture large-scale structures in complex, asymmetric flows, but they also underscore the need for robust data processing. Here, we present a BOS dataset for a 3D flight body in supersonic flow, acquired using a fixed seven-camera array positioned around the test section of a high-speed tunnel. The strut-mounted model was rotated through ten angular positions, yielding 70 views over $120^\circ$ of rotation and enabling detailed estimation of asymmetric features such as an ellipsoidal shock wave that becomes detached at the top. Sample reconstructions are performed with our NIRT algorithm, and neural data assimilation is used to recover additional fields like velocity. The dataset is intended as a community resource for advancing BOS tomography and includes computer-aided designs, calibration and BOS images, masks, deflection fields, reconstructed flow fields, and NIRT code for reproducibility. With these materials, researchers can benchmark current methods, develop new workflows, and evaluate BOS tomography for high-speed flows over realistic flight bodies.\par

\section{Experimental and Computational Methods}
\label{sec: methods}

\subsection{Model Flight Body}
\label{sec: methods: design}
\begin{figure}[ht]
    \centering
    \includegraphics[width=6.5in]{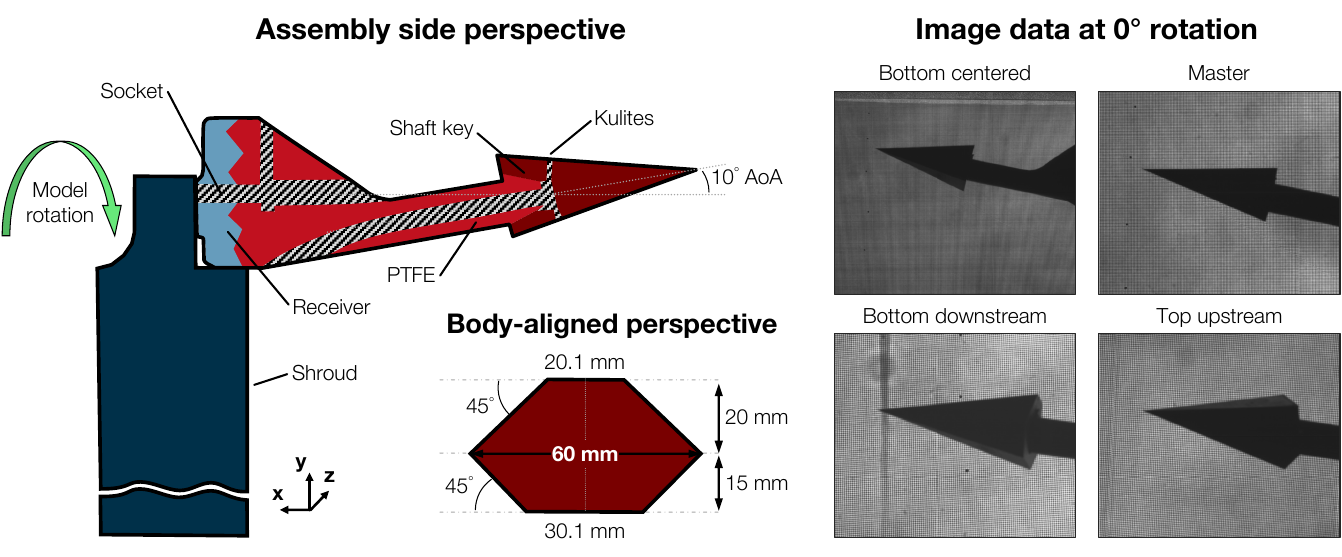}
    \caption{Flight vehicle model with fastening and routing regions shown in zebra print and individual design pieces in unique colors (left). Body dimensions are given (bottom) alongside images of the installation from four perspectives (right).}
    \label{fig: model}
\end{figure}

The model used in this work was a generic flight body set at a positive angle of attack (AoA). It was not intended to represent a real vehicle but rather to produce characteristic flow asymmetries. The design was an asymmetric square-pyramid with a semi-angle of $20.1^\circ$ and a length of 82~mm, tested at an AoA of $10^\circ$. The top and bottom of the body were cut to plane surfaces of $13.7^\circ$ and $10.4^\circ$, respectively. The model, key geometric parameters, and four views of the installation are shown in Fig.~\ref{fig: model}.\par

The body was fastened to the vehicle shaft using male--female dovetail joinery, with glue applied to the coupling for added stability. At the base of the shaft, a circular male-toothed index plate engaged a female receiver, which was secured to a steel support rod using a 10-32 bolt. This indexing system permitted $360^\circ$ rotation in $10^\circ$ increments while maintaining a constant AoA. The shaft and base were fastened to the receiver with a washer and 1/4-20 bolt inserted through the shaft housing. A custom shroud, friction-fitted to the strut beneath the receiver, minimized tunnel blockage and was held in place with the same 10-32 bolt. The assembled strut was installed in the floor mount using a keyless bushing, and the mounting plate remained in place for the duration of testing. Re-indexing to a new angular position involved removing the tunnel window on the camera side, loosening the fastening bolt, rotating to the desired setting, re-torquing, and putting the window back in place.\par

The designs were additively manufactured in polycarbonate using an UltiMaker printer. Polycarbonate was selected to limit model deflection during testing. The peak deflection was estimated to be 1~mm; although this was a minor shift, a stiffer metal construction would be preferable. In practice, the deflection was consistent across runs and did not compromise our tomographic analysis. The vehicle shaft contained an internal void (Fig.~\ref{fig: model}) through which PTFE tubing routed data and power lines from Kulite transducers that were embedded on the leeward and windward faces, carrying them through the body and support, out of the tunnel. The Kulites were sampled at 200~kHz and processed with a 100~kHz filter.\par

\subsection{Testing Conditions}
\label{sec: methods: tests}
Experimental testing was conducted at VKF Wind Tunnel D at AEDC: a pressure--vacuum blowdown facility. High-pressure air was brought to the desired stagnation conditions, routed through a variable-Mach nozzle and square 12-inch test section, and then exhausted into a vacuum sphere. The tunnel can operate from Mach~1.5 to 5 and provide stagnation pressures of 5 to 60~psia. Further details about the facility are reported by Hofferth and Ogg \cite{Hofferth2019}. For this campaign, the nozzle was set to generate Mach~4.8 flow, and runs were conducted at $p_\mathrm{0} = 60$~psia and $T_\mathrm{0} = 120$~$^\circ$F. Table~\ref{tab:tests} summarizes the mean and standard deviation of freestream conditions in the tunnel.\par

\begin{table}[ht]
    \caption{Wind Tunnel Operating Conditions for BOS Tomography Measurements}
    \centering
    \tabcolsep=0.2cm
    \vspace*{.1em}
    \begin{tabular}{c c c c c c c c}
        \hline\hline \\[-.75em]
        \multicolumn{1}{c}{} &
        \multicolumn{1}{c}{\bf $\boldsymbol{p_\infty}$,~Pa} &
        \multicolumn{1}{c}{\bf $\boldsymbol{T_\infty}$,~K} &
        \multicolumn{1}{c}{\bf $\boldsymbol{\rho_\infty}$,~kg/m$\boldsymbol{^3}$} &
        \multicolumn{1}{c}{\bf $\boldsymbol{a_\infty}$,~m/s} &
        \multicolumn{1}{c}{\bf $\boldsymbol{u_\infty}$,~m/s} &
        \multicolumn{1}{c}{\bf $\boldsymbol{\mu_\infty}$,~Pa~s} &
        \multicolumn{1}{c}{\bf Re$\boldsymbol{_\infty}$,~m$\boldsymbol{^{-1}}$} \\ 
        \hline 
        \multirow{1}{*}{}  &  &  &  &\\[-.75em]
        \multirow{1}{*}{} Mean  & 1039.16 & 57.80 & $5.84\times10^{-2}$ & 152.06 & 732.28 & $3.80\times10^{-6}$ & $11.17\times10^{6}$\\
        \multirow{1}{*}{} Std. Dev. & 99.93 & 0.91 & $1.09\times10^{-3}$ & 0.94 & 3.63 & $6.12\times10^{-8}$ & $0.16\times10^{6}$ \\
        \hline\hline
    \end{tabular}
    \label{tab:tests}
\end{table}

Runs began by pulling the facility to vacuum, at which point reference flow-off images were acquired. The tunnel was then brought to the desired condition. Once steady flow was established, flow-on images were taken. This continued until imaging was complete, after which the tunnel was returned to standby conditions. On average, the tests lasted about 200 seconds.\par

\subsection{Experimental Setup}
\label{sec: methods: imaging}

\subsubsection{Camera and Illumination Equipment}
\label{sec: methods: imaging: setup}
\begin{figure}[ht]
    \centering
    \includegraphics[width=6in]{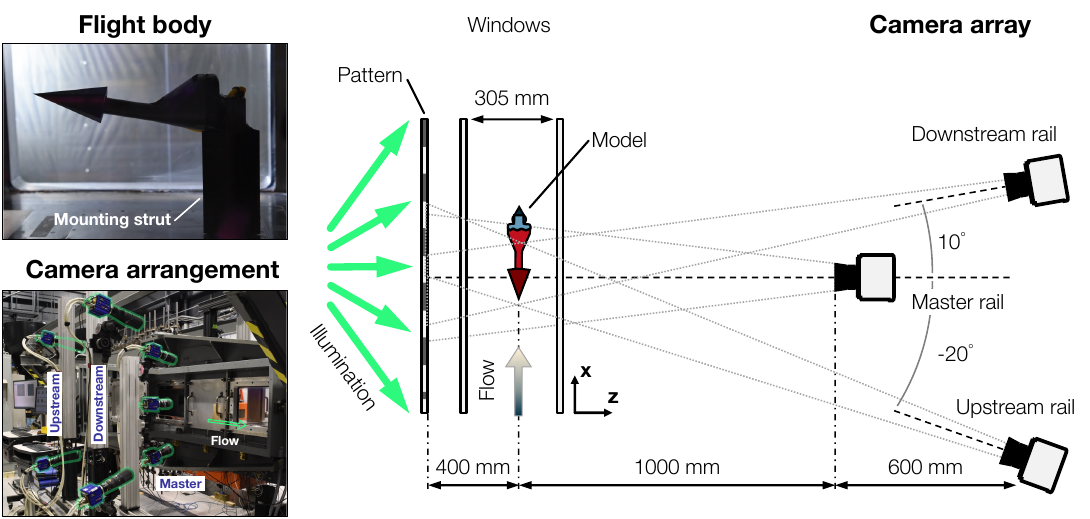}
    \caption{Images of the flight body (top left) and experimental setup (bottom left). Schematic of the imaging setup and facility (right). Cameras are positioned to achieve similar fields of view and circles of confusion at the object plane, with the latter having an estimated diameter of 0.95~mm.}
    \label{fig: setup}
\end{figure}

Imaging was performed using seven cameras and a backlit BOS pattern, as shown in Fig.~\ref{fig: setup}. Monochromatic LaVision sCMOS CLHS Imagers were employed, each having a $2560\times2160$-pixel sensor with 6.5~$\upmu$m pixels. The cameras were mounted on three X95 optical rails: one located 1~m from the model and the other two positioned 1.6~m away, with an upstream rail ($-20^\circ$) and a downstream one ($10^\circ$). A backlit BOS pattern, described below, was mounted 400~mm behind the model. The first rail carried two cameras with 105~mm Nikon lenses and a third with an 85~mm Rokinon lens. The master camera viewed the test article head-on, with a field of view of $158 \times 13$~mm$^2$; the other two cameras on this rail were vertically offset by approximately $\pm15^\circ$. Each of the upstream and downstream rails carried two cameras with 200~mm Nikon lenses, also mounted at $\pm15^\circ$ elevation. The 105~mm and 200~mm lenses were operated at $f/32$, and the 85~mm lens was stopped at $f/22$. Lens-mount adapters were used on all cameras.\par

The cameras imaged the BOS pattern through a pair of 25.4~mm thick sidewall windows, with the model positioned between them. The background panel consisted of an acrylic sheet with a 2D sinusoidal pattern (frequency of 0.8~mm$^{-1}$) printed on a transparency and affixed to the surface; diffusing sheets were placed behind the pattern to homogenize the back-lighting. Illumination was provided by a Spectra-Physics Quanta Ray Nd:YAG laser, delivering 10~ns pulses at 400~mJ and 10~Hz. The laser beam was expanded to illuminate the entire pattern. For each test, 2000 flow-off and flow-on images were acquired at 10~Hz with 15~$\upmu$s exposures, freezing the flow within each pulse (bulk displacement of ${\sim}0.11$~px). Across ten tests, the model was rotated from 0$^\circ$ to 90$^\circ$ in 10$^\circ$ increments, yielding 70 angular views.\par

\subsubsection{Camera Calibration}
\label{sec: methods: imaging: calibration}
Cameras were calibrated using a two-level LaVision 106-10-2 plate, with the background pose estimated from images of a QR5-306-11.6 plate, temporarily mounted in front of the pattern during setup. The two-level plate was fixed to the same strut as the model to aid centerline alignment and it was front-lit for visibility using a pair of floodlights. The calibration setup was rotated in $10^\circ$ increments until the dots were no longer visible. The target was then re-centered and the process repeated in the opposite direction, followed by a final image taken at $0^\circ$ for validation. Thirteen plate positions were used for view registration.\par

Calibration proceeded in two stages. First, the images were preprocessed using an in-house script to detect control points (dots), and each camera was calibrated independently using Zhang’s method \cite{Zhang2002}. Because the reconstruction requires all cameras to share a common global coordinate system, the per-camera extrinsics from this stage were transformed into a global frame defined by a calibration target visible to all cameras. The second stage refined this registration using an in-house nonlinear optimization. The cost function was the mean reprojection error across all cameras, and the control vector included the extrinsics of every camera together with those of all calibration target poses except the global reference, so that one target pose fixed the frame. Hence, the optimization enforced mutually consistent projective geometries. It was performed using MATLAB's trust-region-reflective least-squares algorithm in \texttt{lsqnonlin}. Afterwards, all cameras were shifted to the model's rotation axis. A $-4.75$~mm translation in the $z$-direction, a $1$~mm offset in $y$, and a $0.5^\circ$ clockwise rotation were applied to the global frame align it with the model.\par

The Perspective-n-Point (PnP) algorithm was used to estimate the background pose, given that only a single measurement plane was available and only few dots were visible \cite{Gao2003}. Known camera intrinsics were used to establish correspondences between 3D dot locations and their 2D image projections, yielding the position and orientation of the background. The calculation was done with MATLAB's \texttt{estworldpose} function \cite{ComVisToolbox}; the resulting pose was refined with an in-house optimization and transformed to the same global coordinate system as the cameras.\par

Average root-mean-squared reprojection errors for the camera and background calibrations were 0.46~px and 0.49~px, respectively, both below half a pixel. The corresponding standard deviations were 0.22~px and 0.25~px, all of which are typical values for experimental calibrations in ground-test facilities. Camera calibration errors in these tests likely stem from dot detection errors caused by suboptimal illumination at extreme angles; limited optical access and small apertures exacerbate this issue. For the background pose, the choice of PnP hyperparameters (inlier threshold, maximum iteration count) can limit accuracy with sparse point correspondences.\par

\subsection{Data Analysis}
\label{sec: methods: analysis}

\subsubsection{Deflection Sensing}
\label{sec: methods: analysis: deflection sensing}
\begin{figure}[ht]
    \centering
    \includegraphics[width=6.5in]{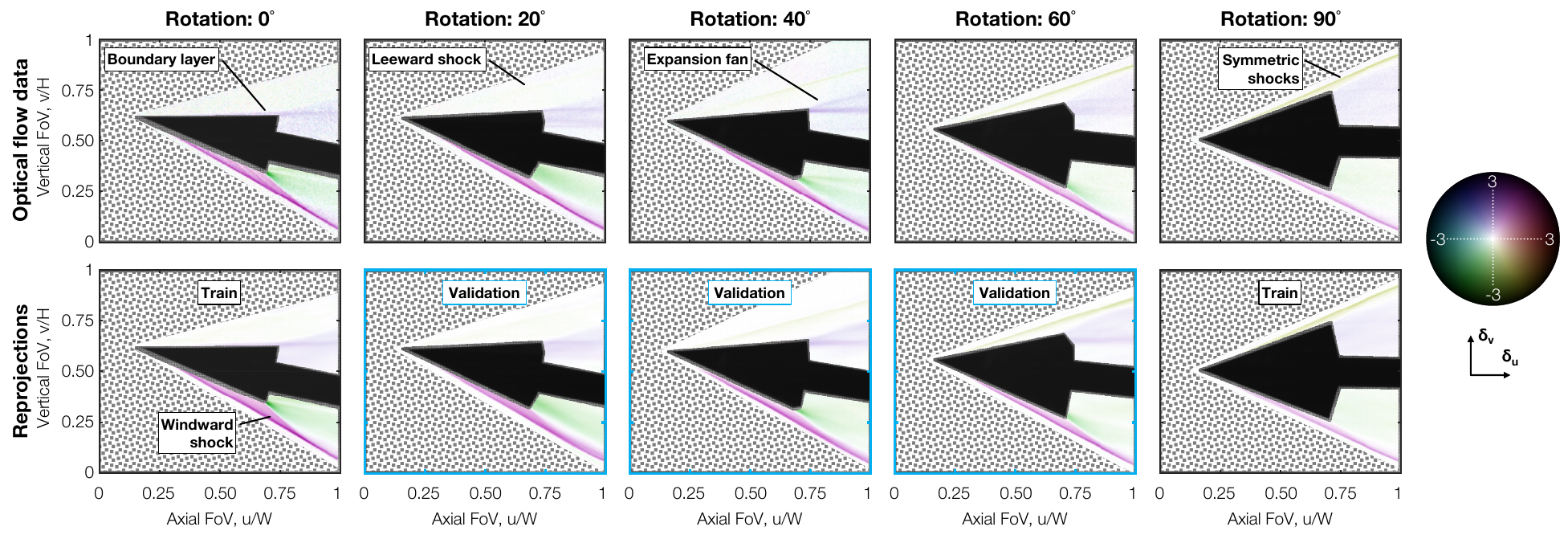}
    \caption{Deflections from the master camera obtained by optical flow (top row) and reprojected through the reconstructed 3D field (bottom row). Blue borders indicate validation frames excluded from reconstruction. Deflections are color-coded by the bi-directional wheel (right), with white indicating no deflection. Masks used in reconstruction are shown: grayscale regions mark omitted pixels; the dotted pattern marks the freestream mask, where an ambient density boundary condition was weakly enforced.}
    \label{fig: deflections}
\end{figure}

Computer vision algorithms were used to estimate bi-directional deflections from the flow-off\slash flow-on image pairs \cite{Schmidt2025}. Figure~\ref{fig: deflections} shows data from the master camera at different rotations of the flight body. Deflections estimated via Horn--Schunck optical flow are presented in the top row, with the bottom row showing reprojections computed from the results in Sec.~\ref{sec: results}. Characteristic shock structures are evident, including the expected windward and leeward oblique shocks and a clear expansion fan forming off the tail. reprojections outlined in blue correspond to views not included in the reconstruction and thus serve as validation, corroborating the accuracy of our results.

Our Horn--Schunck implementation includes local image normalization within 16~px windows and Gaussian smoothing using a 1~px kernel, applied to both the reference and deflected frames. Optimal regularization was approximated through an L-curve analysis. This method produced estimates consistent with physical intuition, in contrast to tuned cross-correlation (CC) and wavelet optical flow (WOF) estimates. The CC scheme underestimated deflections across shocks and exhibited greater artifacts than the Horn--Schunck estimates shown in Fig.~\ref{fig: deflections}. WOF results had magnitudes comparable to Horn--Schunck, but WOF deflections displayed wavelet ``fingerprints,'' potentially caused by spatial variations in the back-illumination. Visual comparisons are available in the public repository.\par

Three masks were generated for each view: model, ambient, and active masks. Model masks were manually created using the mask drawing functionality in \texttt{PIVlab} \cite{Stamhuis2014}, they included any pixel occupied by the body in either the flow-off or flow-on images for a given view. In some frames, e.g., the $0^\circ$ rotation in Fig.~\ref{fig: deflections}, the background pattern remained visible because the model deformed by up to 1~mm in the $y$-direction under aerodynamic loading. Pixels in the model mask were omitted from the measurement loss, although the volume up to the deformed model surface remained included in the regularization terms for tomography and data assimilation. Ambient masks were created in a similar fashion, with freestream regions identified visually based on the observed shock structures in the deflection fields; extra care was taken in views with weak signals, such as the $40^\circ$ rotation in Fig.~\ref{fig: deflections}. Deflections in these ambient regions were centered about zero with standard deviations of order 0.1~px, roughly $4\%$ of the deflection magnitude across shocks and expansion fans. Hence, these regions were treated as freestream for mean flow reconstructions, as described in \cite{Molnar2025}. Lastly, the active mask was defined as the complement of the union of the model and ambient masks.\par

\subsubsection{Tomography and Data Assimilation}
\label{sec: methods: analysis: reconstruction}
A NIRT framework was used for tomographic reconstruction as well as data assimilation \cite{Molnar2025}. Networks comprised six layers with 300 nodes each, implemented in TensorFlow~2.10.1, and symmetry about the centerline was weakly enforced. Fourier encodings with 256 features were drawn from a normal distribution with a standard deviation of 16~m$^{-1}$. Weights were initialized from a standard normal distribution and biases set to zero. Integrals in the loss terms were approximated by Monte Carlo sampling. Regularization and boundary losses were evaluated using batches of 10~000 points. The data loss used batches of 500 pixels, with synthetic measurements rendered by a cone--ray operator. The operator's coefficient of variation was 8.5\%, requiring 8000 points per pixel. Sampling for the data loss was restricted to a conical hull around the flight body to reduce computational costs.\par

Total variation (TV) regularization was used for tomography to accurately recover sharp-edged structures like shock waves. For data assimilation, the TV term was replaced by residuals from the Euler equations. Nine views from the master-camera column were assimilated, corresponding to model rotations of $0^\circ$, $50^\circ$, and $90^\circ$. Although the trained networks provide continuous estimates of flow fields, we plot them using a voxelized grid, with $400\times350\times350$ voxels in a $150\times130\times130$~mm$^3$ volume (375~$\upmu$m voxel side length).\par

\section{Results}
\label{sec: results}
\begin{figure}[ht]
    \centering
    \includegraphics[width=6.5in]{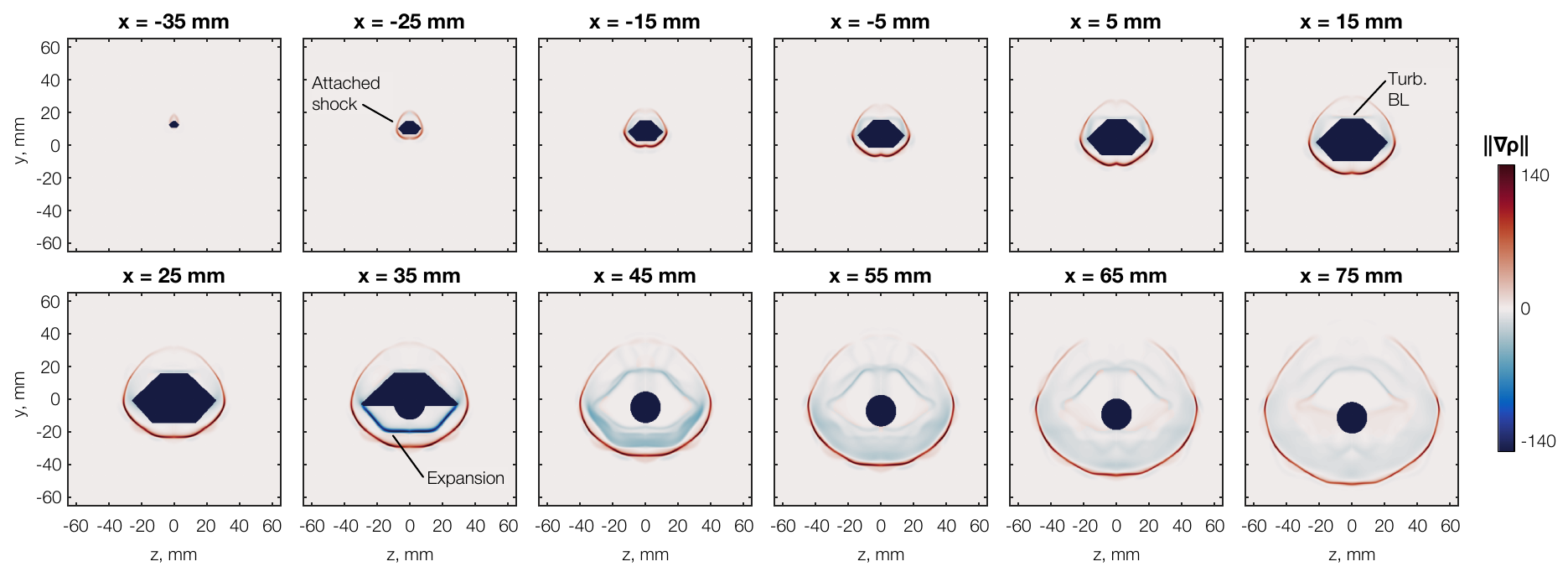}
    \caption{Streamwise slices of the non-dimensional density gradient field with key flow features labeled. Magnitudes are colored by the sign of the streamwise density gradient to distinguish expansion (blue) from compression (red).}
    \label{fig: gradients}
\end{figure}

We first apply NIRT with TV regularization. Figure~\ref{fig: gradients} shows slices of the non-dimensional density gradient field from the reconstruction, progressing in the streamwise direction. Attached oblique shocks, resembling a conical shock wave, are visible at $x = -25$~mm. Further downstream, at $x = 15$~mm, boundary layer growth along the top of the model is well captured and indicated in blue, where body-normal gradients dominate. The boundary layer thickness is consistent with a fully turbulent profile. An expansion fan emanating from the tail is highlighted at $x = 35$~mm and persists in the downstream frames. This structure matches the flight geometry well and exhibits minimal reconstruction artifacts. The apparent curling of the windward oblique shock near the centerline at mid-span (clearly visible at $x = 5$~mm, e.g.) may stem from the limited view count, calibration errors, or noise. Similarly, the rippling expansions near the body (low-magnitude blue oscillations) are most likely artifacts. Even so, the results represent an improvement over prior work, where errors such as non-physical expansions ahead of shocks and streaks in the flow were more pronounced.\par

\begin{figure}[ht]
    \centering
    \includegraphics[width=3.5in]{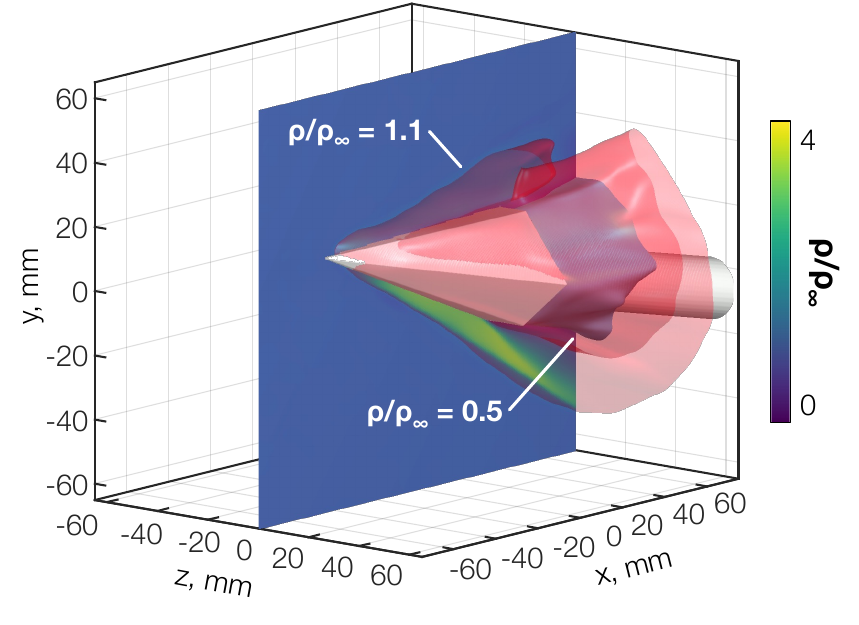}
    \caption{Density field shown as a centerline slice and two isosurfaces. The higher level ($\rho/\rho_\infty = 1.1$, red surface) captures compression structures as the flow encounters the body; the lower level ($\rho/\rho_\infty = 0.5$, blue surface) depicts the expansion off the tail.}
    \label{fig: surface}
\end{figure}

Next, we highlight 3D flow structures in Fig.~\ref{fig: surface} by visualizing the density field in the $y$--$z$ plane together with two isopycnic surfaces, which indicate a compression and expansion. Oblique shocks appear as a red surface, with variation across the surface indicated by the slice's green--yellow gradient. Expansion off the tail is also evident as the blue surface, with a profile that hews closely to the body shape. Freestream artifacts are minimal, underscoring the resilience of our approach to noise and deflection sensing errors. During testing, the model exhibited repeatable elastic deflection up to 1~mm in the $y$-direction, which interfered with deflection sensing near the tip. Despite this, the reconstruction remains physically consistent along the body, with only a thin sliver of the nose protruding through the compression isosurface at the selected isolevel.\par

\begin{figure}[ht!]
    \centering
    \includegraphics[width=6.5in]{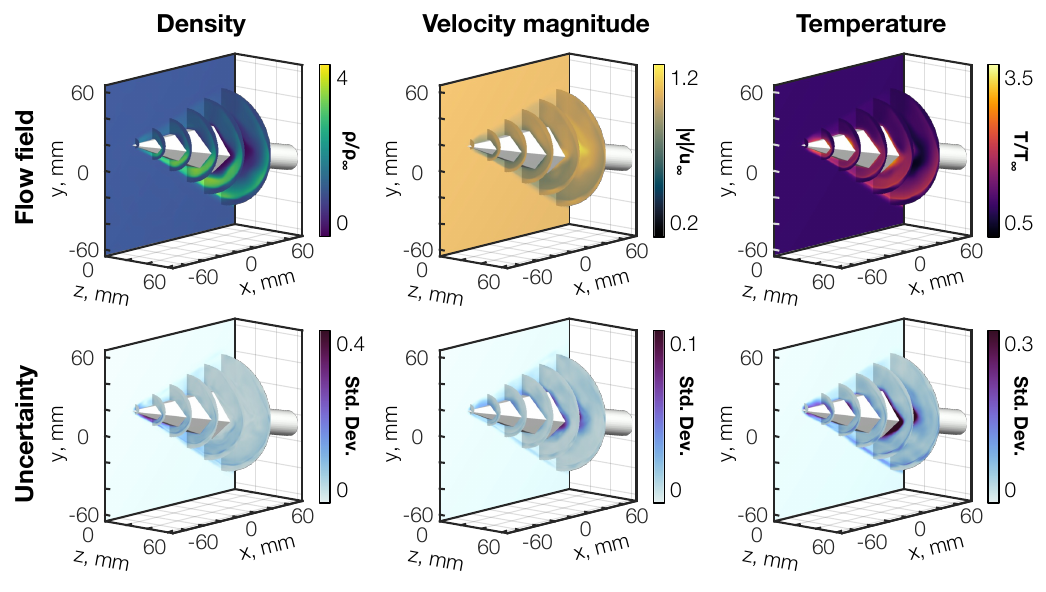}
    \caption{Flow fields (top) and uncertainty fields (bottom) from Euler-based data assimilation in the NIRT framework. Steep gradients through the shock are resolved across all fields with minimal artifacts.}
    \label{fig: panel}
\end{figure}

Lastly, we present data assimilation results in Fig.~\ref{fig: panel}, showing slices of field variables, normalized by the values in Table~\ref{tab:tests}, positioned atop their associated uncertainties. Uncertainty estimates combine aleatoric and epistemic components: aleatoric uncertainty reflects run-to-run variations, while epistemic uncertainty is derived from an ensemble of ten reconstructions initialized with different seeds. Total uncertainty is computed via a first-order Taylor expansion under the assumption of constant Mach number. This estimate accounts only for freestream variability and NIRT-related uncertainty, excluding contributions from calibration or deflection sensing. The density field closely matches the TV results, though compression near the tip is sharper. Penalizing Euler residuals improves infilling in regions with minimal image data but also increases uncertainty there. Unmeasured quantities like velocity and temperature are effectively recovered. As expected, the flow decelerates through the compression over the flight body, then quickly accelerates during the expansion off the aft. Uncertainty patterns align with physical intuition, concentrating along shock fronts, where calibration, deflection sensing, and reconstruction errors compound, and in the expansion, where signal levels drop off. These trends are consistent with the sensitivity of shock angles to small changes in run conditions. The inferred temperature field is especially notable given the difficulty of experimentally resolving near-wall temperature gradients.\par

\section{Conclusions}
\label{sec: conclusions}
This work establishes an open-source BOS tomography database for high-speed flow over a flight body, containing all the components needed for algorithm development and benchmarking. The release includes computer-aided designs, image data (calibration, flow-off, flow-on, masks), deflection estimates, 3D reconstructions, and the full tomography codebase. Using our NIRT framework, we analyze a representative subset of the data and present results from each stage of processing. Among the tested deflection sensing algorithms, Horn--Schunck optical flow with local image normalization performs best. Reconstructions capture steep shock fronts across the domain, an expansion fan consistent with the model geometry, and exhibit minimal artifacts. Finally, we demonstrate, for the first time, 3D data assimilation directly from schlieren measurements, using the Euler equations to refine the density field while simultaneously recovering the velocity and temperature fields.\par

\subsection*{Data Availability}
The full dataset, including component designs, calibration images, flow-off\slash flow-on image pairs, masks, deflections, and 3D reconstructions, along with the camera calibration, deflection sensing, and NIRT code, are available at \doi{10.26208/1VE2-5C19}.\par

\subsection*{Acknowledgments}
All authors acknowledge the support of the Wind Tunnel D operating crew, including Mr.~Phil Vernon, Mr.~Jacob Floyd, Mr.~Ben Jonakin, and Ms.~Britt Osborne. J.P.M. acknowledges support from a DoD NDSEG Fellowship. S.J.G. acknowledges support from NSF under Grant No.~2227763.\par


\end{document}